\documentclass[prd, twocolumn, nofootinbib, floatfix]{revtex4-1}
\usepackage{amsmath}
\usepackage{graphicx}
\usepackage{dcolumn}
\usepackage{bm}
\usepackage{epsfig}
\usepackage{amssymb,latexsym,mathrsfs}
\usepackage{graphicx}
\usepackage{color}
\usepackage{hyperref}

\hypersetup{
    colorlinks=true,
    linkcolor=red,
    citecolor=blue,
} 

\newcommand{\be}{\begin{equation}}
\newcommand{\ee}{\end{equation}}
\newcommand{\bs}{\begin{split}} 
\newcommand{\bea}{\begin{eqnarray}}
\newcommand{\eea}{\end{eqnarray}}

\newcommand{\om}{\Omega_m}

\newcommand{\bh}{\bar H}

\begin{document}

\title{How Fabulous Is Fab 5 Cosmology?} 
\author{Eric V.\ Linder} 
\affiliation{Berkeley Center for Cosmological Physics \& Berkeley Lab, 
University of California, Berkeley, CA 94720, USA} 

\begin{abstract} 
Extended gravity origins for cosmic acceleration can solve some fine tuning 
issues and have useful characteristics, but generally have little to say 
regarding the cosmological constant problem.  Fab 5 gravity can be ghost free 
and stable, have attractor solutions in the past and future, and possess 
self tuning that solves the original cosmological constant problem.  Here 
we show however it does not possess all these qualities at the same time.  
We also demonstrate that the self tuning is so powerful that it not only 
cancels the cosmological constant but also all other energy density, and 
we derive the scalings of its approach to a renormalized de Sitter cosmology.  
While this strong cancellation is bad for the late universe, it greatly eases 
early universe inflation. 
\end{abstract}

\date{\today} 

\maketitle

%%%%%%%%%%%%%%%%%%%%%%%%%%%%%%%%%%%%%%%%%%%%%%%%%%%
\section{Introduction} \label{Sec:intro} 

Accelerated expansion is a property of our universe of immense importance 
yet lacking a clear physical explanation.  Adding an energy density with 
accompanying negative pressure involves a seemingly arbitrary scalar field 
potential (flat in the case of a cosmological constant) that is moreover 
unnatural, not just in initial conditions but in its reaction under quantum 
radiative corrections.  This last aspect can be partly ameliorated by 
imposing, e.g., a shift symmetry on the field but much arbitrariness 
remains.  One approach is to remove the potential entirely and deal with 
only the kinetic behavior of the field, as in k-essence 
\cite{armend1,armend2,chiba,07050400}. 

Altering the gravitational theory is another approach.  Within scalar-tensor 
theories one can again employ shift symmetric fields, and considerable work 
has been done on these Galileon theories \cite{nicolis,deffayet1,deffayet2}.  
None of these address why the vacuum energy does not generate a cosmological 
constant characteristic of the Planck energy scale or other early universe 
scale.  One interesting development has been the identification of four 
unique self tuning terms in the action, called the Fab Four 
\cite{fab4a,fab4b,fab4c}, that can 
dynamically cancel a high energy cosmological constant. 

An early approach to combining several of these characteristics was the 
purely kinetic coupled gravity of \cite{gubitosi}, using what later became 
called the $L_{\rm John}$ term of Fab Four, but this version was later found 
to be unhealthy due to having a ghost (negative kinetic energy state).  
By generalizing the coefficient of the coupling to a free constant, the 
ghost could be exorcised and this becomes 
the derivatively coupled Galileon model, studied in detail in \cite{appgal}. 
Promoting the constant coefficient to a potential depending on the field 
value leads to the full $L_{\rm John}$ term, and allowing it to depend 
further on the field kinetic term gives the $L_5$ term of Horndeski 
scalar-tensor gravity \cite{horndeski,deffayet1,l5form}.  However, these 
terms by themselves can (and without a potential do) have 
a gradient instability in the early universe, rapidly violating homogeneity. 

By taking the nonlinear generalization of purely kinetic coupled gravity, 
\cite{fab5} showed that one could go beyond Fab Four for obtaining self 
tuning, introducing the name Fab 5 gravity, while avoiding a potential. 
Interestingly, despite the nonlinearity this theory added no new 
propagating degrees of freedom (on a homogeneous, isotropic spacetime) 
and hence avoided a ubiquitous ghost.  Fab 5 gravity was shift 
symmetric and so avoided many naturalness problems with radiative corrections, 
could have an attractor behavior in the early universe and so not only 
removed an arbitrary potential but also fine tuning of initial conditions, 
could have an attractor behavior in the late universe giving cosmic 
acceleration and a de Sitter state, and could dynamically cancel a 
cosmological constant.  It is also related to multifield Galileons 
\cite{12104026}. 

While Fab 5 gravity possesses many interesting and desirable characteristics, 
not the least being cancellation of the high energy cosmological constant, 
\cite{fab5} did not show that these characteristics existed simultaneously. 
In fact, we will find that some are exclusive.  Section~\ref{sec:ghost} 
investigates the health and naturalness of Fab 5 gravity.  We present 
explicit demonstrations of self tuning, including in the presence of other 
energy-momentum components, in Sec.~\ref{sec:tune}.  Conclusions about the 
overall state of the theory are discussed in Sec.~\ref{sec:concl}.

%%%%%%%%%%%%%%%%%%%%%%%%%%%%%%%%%%%%%%%%%%%%%%%%%%%%%%%%% 
\section{Soundness and Naturalness} \label{sec:ghost} 

Fab 5 gravity uses a derivative coupling of a scalar field to the Einstein 
tensor, turning the kinetic term into a disformal theory.  (For more on 
disformal theories, see e.g.\ 
\cite{disf1,disf2,disf3,disf4,disf5,disf6,disf7}.)  
The field has no potential term so the theory is manifestly shift symmetric.  
Matter is minimally coupled.  The derivatively coupled term is promoted to 
a nonlinear function, in a manner similar to how the Ricci scalar is 
promoted to a nonlinear function in $f(R)$ gravity.  This can be viewed as 
creating a new, auxiliary scalar field, but due to the symmetries of the 
Einstein tensor in a homogeneous, isotropic spacetime the new field has no 
propagating degrees of freedom and preserves the second order nature of 
the field equations. 

Explicitly, the action is 
\bea 
S&=&\int d^{4}x\,\sqrt{-g}\left[\frac{M_{\rm Pl}^{2}}{2}\, R+c_{1}X\right. 
\notag\\ 
&\quad\qquad&\left.+f(c_{2}X+\frac{c_{G}}{M^2}G^{\mu\nu}\phi_{\mu}\phi_{\nu})\right] 
+S_m(g_{\mu\nu})\,,\label{eq:act} 
\eea 
where $X=(-1/2)g^{\mu\nu}\phi_\mu\phi_\nu$, $\phi_\mu=\nabla_\mu\phi$, 
$G^{\mu\nu}$ is the Einstein tensor, and $S_m$ is the action for the 
matter fields.  The mass scales are the Planck scale $M_{\rm Pl}$ and 
the scalar field scale $M$, which we normalize to the Hubble constant $H_0$ 
when using Fab 5 to give late universe cosmic acceleration. 

We build on the characteristics of Fab 5 gravity discussed in \cite{fab5}. 
A key ingredient is the nonlinear function $f(\chi)$, where 
\bea 
\chi&=&c_2 X + \frac{c_G}{M^2}G^{\mu\nu}\phi_\mu\phi_\nu\\ 
&=&\frac{c_2}{2}\bh^2x^2+3c_G\bh^4x^2 \ . 
\eea 
We define dimensionless variables $\bh=H/H_0$ 
and $x=d(\phi/M_{\rm Pl})/d\ln a$, and assume a Friedmann-Robertson-Walker 
cosmology.  The major constraints on the health of the 
theory will arise at high redshift, when the $c_G$ term dominates over 
either the $c_1$ or $c_2$ terms (since it has an extra $\bh^2$ factor).  
Therefore our main conclusions are independent of whether we include 
the canonical kinetic term inside the nonlinear function or not, i.e.\ 
whether we set $c_1=0$ or $c_2=0$; for definiteness we set $c_1=0$, 
hence including the usual kinetic term in the function.  This gives a 
disformal structure 
\be 
\chi=\left[\frac{-c_2}{2} g^{\mu\nu}+\frac{c_G}{M^2}G^{\mu\nu}\right] 
\,\phi_\mu\phi_\nu \ . 
\ee

%%%%%%%%%%%%%%%%%%%%%%%%%%%%%%%%%% 
\subsection{General criteria} 

Although the theory is guaranteed by its structure to have second order 
field equations, and that the auxiliary scalar field $\chi$ does not 
propagate and so has no ghost, we still must check that the main scalar 
field $\phi$ is well behaved.  In order to have a healthy theory we must 
require that it is free from ghosts and 
is stable against instabilities.  General expressions for these conditions 
were given in \cite{fab5}; in the early universe limit, where the effective 
energy density contribution $\Omega_\phi\ll1$, these simplified to 
\bea 
3c_G \bh^2\,(f_\chi+2\chi f_{\chi\chi})&>&0 \label{eq:ghost}\\ 
\frac{2\dot H+3H^2}{3H^2}\,\frac{f_\chi}{f_\chi+2\chi f_{\chi\chi}}&>&0 \,, 
\label{eq:stable} 
\eea 
respectively. 

Note that in the linear theory, where $f(\chi)=\chi$ hence $f_{\chi\chi}=0$, 
the stability condition is violated during the radiation domination epoch 
where $\dot H=-2H^2$, 
as pointed out by \cite{appgal}.  This gradient instability (also known as 
Laplace instability) rules out the purely kinetic coupled gravity of 
\cite{gubitosi}.  Furthermore, it constrains derivatively coupled Galileons 
and many Fab Four gravity models to have the derivative coupling and 
$L_{\rm John}$ 
respectively to be unimportant aspects of the theory in the early universe, 
considerably weakening their utility. 

Thus, the nonlinear promotion at the heart of Fab 5 is a key element 
guarding against instability and turning the theory healthy.  The form of 
$f(\chi)$ must be chosen to satisfy stability. 

In addition to the absolute requirements of Eqs.~(\ref{eq:ghost}) and 
(\ref{eq:stable}) to have a healthy theory, there are some desiderata 
that allow it to thrive.  We might ask that the model be reasonably 
natural, i.e.\ not fine tuned in its initial conditions.  (Note that the 
theory has technical naturalness due to shift symmetry, i.e.\ once 
the initial conditions are imposed they will not obtain large quantum 
corrections.  We do, however, have to choose $M=H_0$ in order for 
$c_2,\,c_G\sim{\mathcal O}(1)$ to give current cosmic acceleration, as 
for any theory in the literature.)  Since one of the most interesting 
properties of Fab 5 
gravity is its self tuning, we might ask that the model succeed in tuning 
a large early universe cosmological constant to zero.  Finally we might 
ask that it accords with observations, having the standard progression 
of radiation dominated epoch, matter dominated epoch, and late time 
cosmic acceleration with a matter density contribution today $\om\approx0.3$. 

Note that a model that achieves all of these would be counted as wildly 
successful.  Most cosmic acceleration models are fine tuned and few are 
technically natural (the cosmological constant itself being a notable 
failure).  None can dynamically cancel a large cosmological 
constant.  For example, if we only succeed in the two requirements (stable 
and no ghosts) and the last desideratum (agreement with observations) then 
we have done as well as $f(R)$ theories, with the bonus of adding stability 
to quantum corrections and having an innate de Sitter attractor in the future. 

%%%%%%%%%%%%%%%%%%%%%%%%%%%%%%%%% 
\subsection{Power law models} 

In the case of power law models, $f(\chi)=A\chi^n$, the stability 
condition in the radiation dominated era requires 
\be 
n<1/2 \,, 
\ee 
as noted by \cite{fab5}.  However, in this era the effective dark energy 
density is on an attractor trajectory, scaling with expansion factor as 
\be 
\rho_\phi\sim a^{-2n/(2n-1)} \,, 
\ee 
and so it is phantom, growing with time.  (We require $n>0$, otherwise in 
the Minkowski vacuum $f(0)\ne0$ and we have included an explicit cosmological 
constant.)  In order for the dark energy not to overwhelm 
the standard radiation or matter eras this requires a fine tuning, one 
more extreme than for the cosmological constant (which corresponds to $n=0$). 

Furthermore, the no ghost condition implies 
\be 
A\,(2n-1)>0 \,, 
\ee 
requiring $A<0$.  (We keep $c_G>0$ otherwise at early times $\chi<0$, 
causing problems for the necessarily noninteger $n$.)  The magnitude of 
the dark energy density at early times is 
\be 
\rho_\phi\approx (4\chi f_\chi-f)= A\,(4n-1) \,, 
\ee 
and hence for $1/4<n<1/2$ we have $\rho_\phi<0$.  This is potentially a 
good thing, since if $\rho_\phi>0$ and phantom (as for $0<n<1/4$) the fine 
tuning is strong.  Moreover, for self tuning one requires $\rho_\phi<0$ in 
order to cancel a positive early universe cosmological constant.  Finally, 
although its magnitude initially becomes more negative (being phantom), at 
later times it can turn around and become positive giving the usual cosmic 
acceleration. 

\begin{figure}[htbp!] 
   \centering
\includegraphics[width=\columnwidth]{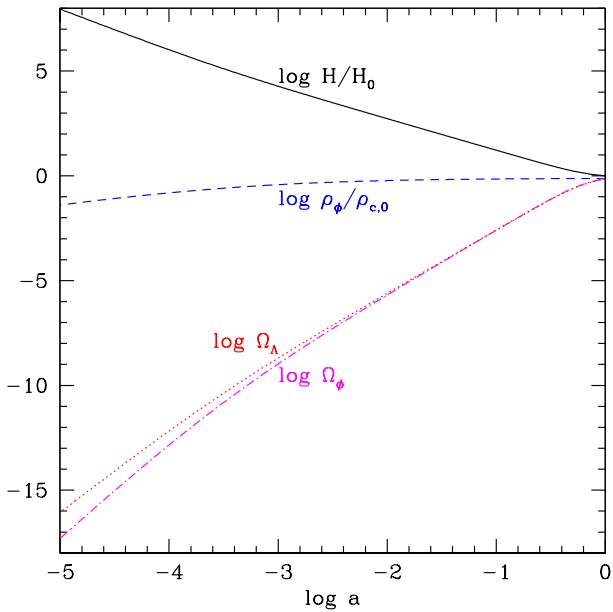} 
\caption{Healthy models are forced to be phantom models at early times, 
so there is a fine tuning in the initial energy density of the scalar 
field.  If we keep the energy density positive, then we require $0<n<1/4$ 
for the early time behavior.  Here we show a case with $n=0.2$ at high 
redshift, which is fine tuned, but not much more than a cosmological constant. 
} 
\label{fig:bpl} 
\end{figure}

Thus for power law models we will always have some fine tuning but may have 
the usual cosmic acceleration or may have self tuning.  First consider the 
least promising case of $0<n<1/4$.  Here the density is positive and 
increases so we must start at a (fine tuned) low level in order not to 
violate radiation domination.  Eventually the density is so strong that 
it dominates, but continues growing and does not approach a de Sitter 
attractor.  Figure~\ref{fig:bpl} shows a broken power law model that does 
go to a de Sitter state.

%%%%%%%%

Next, take $1/4<n<1/2$.  Here the negative dark energy density can self 
tune, canceling a large cosmological constant.  The expansion approaches a 
de Sitter attractor given by $\bh_{\rm dS}=\sqrt{-c_2/(6c_G)}$.  At this 
fixed point $x\to\,$const, i.e.\ the field keeps rolling to dynamically 
cancel the other, positive energy densities, leaving behind a net small 
positive energy density.  That is, Fab 5 turns a large early universe 
cosmological constant into a small late universe cosmological constant, 
but as discussed in Sec.~\ref{sec:tune} does not generally give complete 
radiation and matter dominated eras. 

Within the restrictions of the power law form, we end up with no fully 
satisfactory cosmology, even without self tuning.  It is a general property 
of the power law models that in order for them to be Laplace stable, they 
must also be phantom and hence more fine tuned than a cosmological constant. 
Therefore we now consider non-power law models.

%%%%%%%%%%%%%%%%%%%%%%%%%%%%%%%%%%% 
\subsection{Non-power law models} 

Fab 5 gravity has early time attractor solutions during both radiation 
and matter domination, one of its positive aspects in ameliorating fine 
tuning.  For radiation domination this gives \cite{fab5} 
\be 
\chi\sim a^{-2/(1+2b)} \,, 
\ee 
where $b=\chi f_{\chi\chi}/f_\chi$.  
The no ghost and Laplace stability conditions can be written as 
$f_\chi\,(1+2b)>0$ and $1+2b<0$ respectively, so we see that we must have 
$b<-1/2$ and $f_\chi<0$.  For power law models this gave us $n<1/2$ and 
$A<0$, but now we have more freedom. 

The stability condition assures that $\chi$ grows during radiation 
domination, therefore it starts small, as does the function $f(\chi)$ 
since we require it adds no explicit cosmological constant, so $f(0)=0$. 
Since that is the minimum, we expect that $f$ should look like a power law 
in the early universe.  That is, if the function is analytic near 0 
then we could expand in a Maclaurin series and have 
$f(\chi)\approx f(0)+\chi\,f_\chi(0)+\dots$, or if $f$ is a noninteger 
power law then we reduce to the previous case.  One exception 
is where all the derivatives of $f$ vanish at the origin.  
Another possibility is to allow $f$ to approximate 
a power law but with a constant term so that $\rho\approx(4\chi\,f_\chi-f)$ 
is not proportional to $\chi^n$ and hence does not start so small that it 
is fine tuned.  Finally, we could break the attractor behavior (for example 
by choosing $c_2$ huge compared to $c_G$, so that the derivative coupling 
is unimportant) but this requires fine tuning.  

Considering the first two cases, examples are 
\bea 
{\rm Case\ 1}&:& \qquad f=A\,e^{-\chi_0/\chi} \\ 
{\rm Case\ 2}&:& \qquad f=A\,(e^{-\chi/\chi_0}-1) \ . 
\eea 
Case 1 has $b=\chi_0/\chi-2$; in the early universe when $\chi$ should be 
small then $b$ violates the stability condition.  Case 2 has $b=-\chi/\chi_0$ 
and again is unstable sufficiently early.  We have not been able to find 
any form that both preserves stability and is not fine tuned in the early 
universe (again, recall we are holding ourselves to a high standard -- 
neither the cosmological constant nor $f(R)$ gravity for example satisfy 
the criteria). 

In summary, we have not been able to resolve the tension that avoiding 
instability enhances fine tuning, and avoiding fine tuning leads to 
instability.

%%%%%%%%%%%%%%%%%%%%%%%%%%%%%%%%%%%%%%%%%%%%%%%%%%%%%%%%% 
\section{Canceling the Cosmological Constant} \label{sec:tune} 

\begin{figure}[htbp!] 
   \centering
\includegraphics[width=\columnwidth]{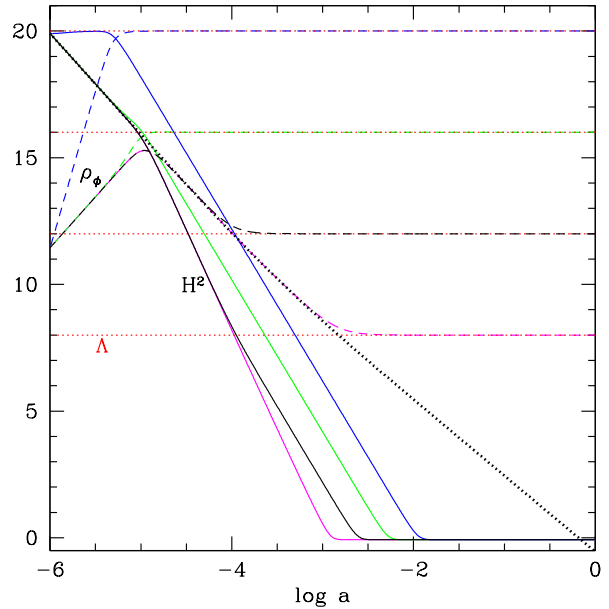} 
\caption{Four cases of large bare cosmological constant 
$\log\rho_\Lambda/(3H_0^2M_{\rm Pl}^2)$ are shown by the horizontal, 
red dotted lines.  The dark energy density 
$\log[-\rho_\phi/(3H_0^2M_{\rm Pl}^2)]$ for power law Fab 5 gravity with 
$n=0.4$ is plotted by dashed lines, with the late time asymptotes 
corresponding to the $\Lambda$ cases.  The field dynamically 
evolves to cancel the bare cosmological constant, regardless of its value. 
The resulting cosmic expansion $\log\bh^2$ is shown by the solid curves 
(of the corresponding colors), 
all asymptoting to a de Sitter state $\bh^2_{\rm dS}=-c_2/(6c_G)$ 
equivalent to a small, late time cosmological constant.  The standard 
radiation and matter expansion history in the absence of Fab 5 is given by 
the dark bold dotted diagonal line. 
} 
\label{fig:powtune} 
\end{figure}

Let us return to perhaps the most interesting feature of Fab 5 (and Fab 
Four) gravity, the possibility of self tuning to cancel a high energy 
scale cosmological constant.  As mentioned in the previous section, and 
illustrated in Fig.~\ref{fig:powtune}, this does indeed work, but works 
too well.  Because the field is coupled to the Einstein tensor 
($G^{\mu\nu}\sim H^2\sim\rho$), the field dynamically cancels 
{\it all\/} energy density, leaving behind only a small cosmological 
constant giving de Sitter expansion with $\bh_{\rm dS}=\sqrt{-c_2/(6c_G)}$.  
It makes no distinction between a large initial cosmological constant and 
other forms of energy density such as radiation and matter. 

Figure~\ref{fig:powtune} exhibits the behavior of the cosmological expansion 
$H^2$ and the dark energy density $\rho_\phi$ for several cases.  We find 
that the self tuning indeed cancels an arbitrarily large cosmological 
constant -- in contrast to \cite{fab5} where the (Laplace unstable) $n=1.5$ 
case had a limited range of self tuning.  However, the self tuning is in 
fact too powerful and does not deliver an observationally viable cosmology.

%%%%%%%%

We see that the resulting expansion history deviates at early times from 
the standard radiation domination and matter domination.  While it follows 
this initially, it begins to deviate once the Fab 5 density has grown in 
amplitude (due to its phantom nature) to become comparable to the background 
energy density $\rho_b$.  Recall that the field evolves along an attractor 
as $\rho\sim a^{-3(1-w_b)n/(2n-1)}$ as long as its energy density is small.  
For the $n=0.4$ case plotted, this corresponds to $a^4$ when $\rho_b$ is 
dominated by radiation ($w_b=1/3$) and $a^{12}$ when $\Lambda$ dominates 
(as in the top, blue case).  

Once $\rho_\phi$ has grown sufficiently, the field ``eats'' the background 
energy density and $\rho_\phi\approx -\rho_b$.  It almost exactly cancels it, 
even through the transition when the background density changes from being 
dominated by radiation and matter to being dominated by the bare $\Lambda$, 
at which point $\rho_\phi\approx -\rho_\Lambda$ from then on.  It leaves 
behind a small, positive, ``renormalized'' cosmological constant 
$\Lambda/(H_0^2 M_{\rm Pl}^2)=-c_2/(6c_G)$.  During the 
time when $\rho_\phi\approx -\rho_r$, the expansion rapidly evolves toward 
the de Sitter attractor $H^2_{\rm dS}$ as a highly negative power of $a$ 
($a^{-8}$ in the case shown in Fig.~\ref{fig:powtune}), switching over to 
$a^{-6}$ when $\rho_\phi\approx -\rho_\Lambda$. 

From the coupled evolution equations for the field $\phi$ and expansion 
rate $H$, given in \cite{fab5}, we can find attractor solutions.  When 
Fab 5 is self tuning in a background otherwise dominated by an energy 
density with equation of state $w_b$, the scalings are 
\bea 
x&\sim& a^{\frac{3}{2}\frac{1+w_b(1-4n)}{n}}\\ 
H&\sim& a^{-\frac{3}{2}\frac{1+w_b(1-2n)}{n}} \ . 
\eea 
In the presence of a large cosmological constant ($w_b=-1$), this yields 
$H^2\sim a^{-6}$ (as noted by \cite{fab4c} for the linear case).  This 
should not be viewed as kination or a stiff fluid ($w_\phi=1$), 
however, as the field velocity $\dot\phi\sim a^3$ is increasing not 
decreasing due to the nonlinear kinetic terms, and the dark energy density 
is constant.  Note that in this case the results are independent of the 
power law index $n$; indeed this attractor is independent of the form of 
$f$ (as long as self tuning operates to achieve the attractor dynamics).  
In the presence of background radiation or matter dominating over a 
cosmological constant, the scaling during self tuning does depend on $n$. 

Previous works on self tuning with Fab Four \cite{fab4a,fab4b,fab4c} and Fab 5 
\cite{fab5} investigated the case with solely the scalar field and a large 
cosmological constant, and so did not emphasize the fact that the standard 
radiation or matter dominated eras are eaten.  As $G^{\mu\nu}\sim H^2$ 
is independent of the form of $f$, there does not seem to be a way around 
this.  Indeed, it does not even depend on nonlinearity of $f$, so 
$L_{\rm John}$ in Fab Four, or its Horndeski theory equivalent, also appear 
in danger if they are dominant at late times.  It is not clear whether 
adding a potential (breaking shift symmetry) or a function of $\phi$ and 
$X$ can ameliorate this (though \cite{fab4c} used a potential to give 
``fake'' radiation and matter eras). 

While this is bad news for self tuning to explain late universe cosmic 
acceleration, it can be very useful for early universe inflation.  The 
self tuning dramatically increases the ease of inflation starting, 
regardless of the other energy density components.  Even for potentials 
that would not normally allow slow roll, the kinetically coupled gravity 
converts them into reasonable inflation models.  This was discussed for 
the linear function $f$ in \cite{12015926}.  The derivative coupling to 
the Einstein tensor has also been used for inflation in 
\cite{09100980,10032635,12034446}.  Indeed the self tuning means that 
inflation can even occur without any potential, since Fab 5 has none.  
The expansion rate during inflation will be $H_{\rm dS}=M\sqrt{-c_2/(6c_G)}$ 
and by choosing the mass scale $M$ to be a high energy scale rather than 
$H_0$ then the values of $c_2$ and $c_G$ remain of order unity.

%%%%%%%%%%%%%%%%%%%%%%%%%%%%%%%%%%%%%%%%%%%%%%%%%%%%%%%%% 
\section{Conclusions} \label{sec:concl} 

The bright hopes for Fab 5 gravity being simultaneously healthy, 
natural, observationally viable, and solving the cosmological constant 
constant problem have been shown to be fabulous in the sense of being 
fictional.  The theory can deliver these characteristics individually 
but not simultaneously.  We emphasize this merely puts the theory on 
the same level as many others considered for dark energy, no worse but 
with equally regrettable aspects. 

These results show the potential danger of coupling fields nonminimally, in 
particular with regard to instabilities, and the strong constraints 
this engenders on the structure of the model.  We find that stable forms 
of the theory require fine tuning of the initial conditions to allow 
radiation and matter domination, while viable cosmological solutions 
with respect to the expansion history tend to have instabilities in the 
field perturbations.  We also note that the derivative coupling affects 
not only the scalar sound speed but also shifts the speed of gravitational 
waves from being the speed of light \cite{fab5}; this may lead to 
gravitational Cherenkov radiation, which is highly constrained \cite{kimura}. 
On solar system scales further considerations arise (though Vainshtein 
screening should remain), as discussed for 
Fab Four by \cite{13105058}; for Fab 5, research is in progress to evaluate 
whether the nonpropagating degree of freedom becomes dynamical there. 

Self tuning, or dynamical 
cancellation, of a high energy cosmological constant is one of the freshest 
and most attractive ideas for solving the cosmological constant problem. 
Unfortunately it is less than even a Pyrrhic victory, but rather suicidal, 
as the kinetic gravity coupling considered here (also known as $L_{\rm John}$ 
in the Fab Four, $L_5$ in Horndeski theory, and even appearing in some 
massive gravity theories \cite{11063312}) cancels all other energy density 
including radiation and matter.  On the plus side this can be used in the 
early universe to greatly ease the onset of inflation.

\acknowledgments 

I thank Stephen Appleby, Antonio de Felice, and Shinji Mukohyama for 
helpful discussions, and the Korea Astronomy and Space Science Institute 
for hospitality.  
This work has been supported in part by the Director, Office of Science, 
Office of High Energy Physics, of the U.S.\ Department of Energy under 
Contract No.\ DE-AC02-05CH11231.

%%%%%%%%%%%%%%%%%%%%%%%%%%%%%%%%%%%%%%%%%%%%%% 

\end{document}